\documentclass[9pt,twocolumn,twoside]{osajnl}

\journal{ol} 

\setboolean{shortarticle}{true}

\title{Experimental verification of phase discontinuities induced scintillation enhancement under weak perturbations}

\author[1]{Han-Tao Wang}
\author[1]{Hua-Jun Zhang}
\author[2]{Lu Zhang}
\author[2]{Hui Lin}
\author[1]{Ming-Yuan Ren}
\author[1,*]{Yu Zhang}

\affil[1]{School of Physics, Harbin Institute of Technology, No.92, Xidazhi Road, Harbin 150001, China}
\affil[2]{Science and Technology on Underwater Test and Control Laboratory, Dalian, Liaoning 116013, China}
\affil[*]{Corresponding author: zhangyuhitphy@163.com}

\begin{abstract}
We verify the existence of scintillation enhancement by measuring the scintillation index of a beam composed of two coherent Gaussian vortex beams with $\pm 1$ topological charges propagating through thermally induced turbulence. Further experimental research based on the reference wave interferometric method demonstrates that this phenomenon is caused by the combined effect of a screw dislocation and an infinitely extended edge dislocation, namely the impact of an anisotropic dislocation. The experimental results indicate that the anisotropic dislocation is more sensitive to weak perturbations than an isotropic screw dislocation and an infinitely extended edge dislocation, which means the anisotropic dislocation has potential for weak perturbation measurement. This phenomenon is instructive in further phase discontinuity research.
\end{abstract}

\setboolean{displaycopyright}{true}

\begin{document}

\maketitle

Phase discontinuities comprised of screw dislocations, infinitely extended edge dislocations and limited edge dislocations are widespread in optical fields \cite{basistiy1995optical}. The conception was initially introduced into optics by Nye and Berry, and then relevant researches have been carried out in theory and experiment \cite{nye1974dislocations}. One of these noteworthy fields is the impacts of phase discontinuities on the evolutionary behaviors of optical fields in turbulence, especially the impacts of screw dislocations \cite{cheng2016propagation}. Several types of beams containing screw dislocations (i.e. optical vortices in optics) have been demonstrated to be more resistant to perturbations than the beams without screw dislocations \cite{2016Vortex,li2017influence,liu2013experimental}. However, it has been found that the scintillation index (SI) for the Laguerre-Gaussian beam LG$^{1}_{0}$ at the vortex axis increases sharply from zero to about unity as the turbulence ``turns on.'' \cite{Aksenov:15} The step response occurs due to the phase singularity, not just the initial intensity distribution of the beam. This phenomenon raises a question of whether phase discontinuities could influence the optical fields under perturbations to present some new characters. Our previous research has verified that the coexistence of screw dislocations with infinitely extended edge dislocations induces scintillation enhancement in theory \cite{wang2021phase}. But the result is limited to the approximation we used in previous research, namely weak turbulence and narrow beam width. To find out whether the phenomenon exists widely in optical fields containing infinitely extended edge dislocations and the cause of the phenomenon, we present the experimental results here.

According to Ref. \cite{wang2021phase}, an optical field consisting of two coherent Gaussian vortex beams with opposite topological charges in different overlapping conditions is the simplicity case to achieve the comparison of three types of phase discontinuity conditions, namely only screw dislocations existing, only infinitely extended edge dislocations existing and both two types of phase dislocations coexisting. And the comparison makes it clear to clarify the cause of scintillation enhancement. Therefore, in this paper, we followed the same way and carried out experiment.

\begin{figure}[ht!]
\centering\includegraphics[width=0.6\linewidth]{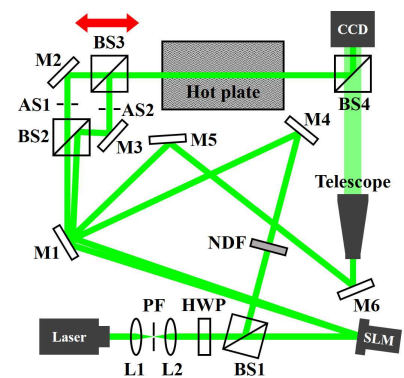}
\caption{Schematic of the experimental setup.}
\label{schematic}
\end{figure}

The schematic of the experimental setup is shown in Fig. \ref{schematic}. A linearly polarized beam generated by a $\lambda = 532 \rm{nm}$ single longitudinal mode laser source is collimated and expanded to a Gaussian beam by a pinhole filter PF and two convex lenses L1 and L2. Then a 50 : 50 beam splitter BS1 divides the Gaussian beam into reference arm and beam shaping arm. In the beam shaping arm, we use a Dammann vortex grating (DVG) with base charge of $l = 1$ to generate several optical vortices with different topological charges \cite{Yu:12}. In order to improve the diffraction efficiency of a spatial light modulator SLM, a half wave plate HWP is used to adjust the polarization of the Gaussian beam. Then the diffraction beam is reflected by a flat mirror M1, and two coherent Gaussian vortex beams in it are separated, selected and recombined using a Mach Zehnder interferometer composed of two flat mirrors M2-3, two aperture stops AS1-2 and two beam splitters BS2-3. It is convenient to change the overlapping degree of two Gaussian vortex beams by adjusting the position of the BS3 in the direction indicated by the red double headed arrow in Fig. \ref{schematic} to realize different combinations of phase discontinuities. The recombined beam propagates through a $20 \rm{cm} \times 30 \rm{cm}$ electric hot plate, which is heated to generate thermal convection to introduce weak perturbations into light transmission, then passes through BS4 and finally arrives at a monochrome charge-coupled device CCD ($964\times1292$ pixels, $3.75 \mu\rm{m}$). The vertical distance from the hot plate to the center of the beam is $2\rm{cm}$. The distance from the right side of the hot plate to CCD is about $0.34\rm{m}$. In the reference arm, the polarized Gaussian beam is reflected by M1 and M4-6 to balance the optical path lengths. A neutral density filter NDF is applied to adjust the intensity of the reference beam to achieve optimal interference, and a telescope aims to expand the reference beam to fully cover the recombined beam. When the interference pattern of the reference beam and the recombined beam is needed to determine the phase distribution of the recombined beam, we let the reference beam propagate through the BS4 to make the interference happen.

\begin{figure}[ht!]
\centering\includegraphics[width=0.75\linewidth]{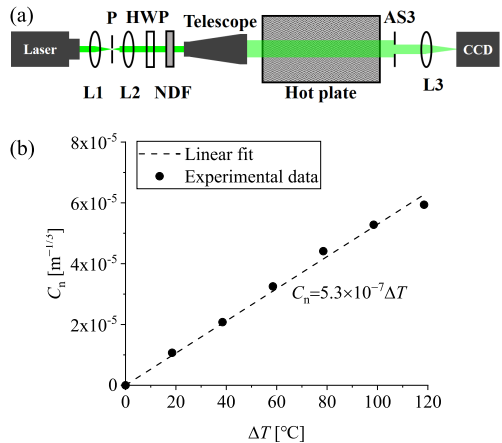}
\caption{(a) Optical setup for angle-of-arrival method and (b) experimentally estimated $C_n$ as a function of $\Delta T$ (in degrees Celsius) using angle-of-arrival fluctuations.}
\label{turbulence}
\end{figure}

Since scintillation is associated with the strength of perturbations (i.e. optical turbulence), the relation between the refractive index structure parameter, $C^2_n$, and the temperature measured by a built-in thermometer of the hot plate should be determined, and the angle-of-arrival method \cite{keskin2006hot,anguita2011influence} was utilized in our experiment. According to the experimental research \cite{anguita2011influence}, we serve the expanded reference beam as the measuring beam, which is modeled as a spherical wave at the receiving plane, and assume the refractive index fluctuations has a Kolmogorov spectrum. The relation between the variance of the angle of arrival $\left\langle {\sigma}^2_{\beta}\right\rangle$ and $C^2_n$ is written as \cite{anguita2011influence}
\begin{equation}
C^2_n = 0.913\left\langle {\sigma}^2_{\beta} \right\rangle 
{D^{{1 \mathord{\left/{\vphantom {1 3}} \right.\kern-\nulldelimiterspace} 3}}}z^{-1},
\end{equation}
where $\left\langle  \cdot  \right\rangle$ represents the ensemble average, $D$ is the aperture diameter of the receiver and $z$ is propagation length of the measuring beam. The experimental setup for angle-of-arrival method and the results are presented in Fig. \ref{turbulence}. The propagation distance $z=1.22\rm{m}$, $D=4.4\rm{mm}$ and the focal length of L3 $f=0.1\rm{m}$ in Fig. \ref{turbulence}(a). Each experimental group of temperature difference $\Delta T$, namely the difference between room temperature $21.5^{\circ}\rm{C}$ and the temperature of hot plate, contains $500$ pictures and the sampling rate was kept as $30\rm{Hz}$. Based on these experimental data, estimated $C_n$ as a function of $\Delta T$ is plotted in Fig. \ref{turbulence}(b). These data fit well with the linear model, which is consistent with the results in Ref. \cite{anguita2011influence}. Thus, it is possible to investigate the scintillation enhancement under weak perturbations with known strength. 

\begin{figure}[ht!]
\centering\includegraphics[width=\linewidth]{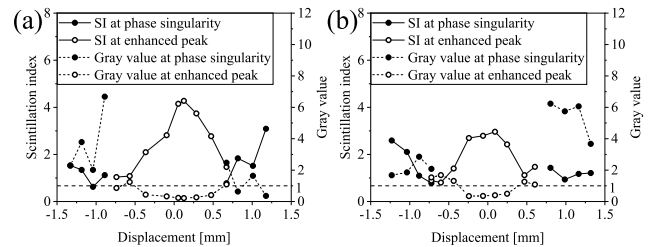}
\caption{Experimental results of the scintillation indexes and the gray values of the pixels at phase singularities and at the peaks of scintillation enhancement versus the overlapping degree of two coherent Gaussian vortex beams (a) when $\Delta T=0^{\circ}\rm{C}$ and (b) when $\Delta T=28.5^{\circ}\rm{C}$.}
\label{comparison}
\end{figure}

However, before studying the relation between turbulence and scintillation enhancement, since the propagating paths of two coherent Gaussian vortex beams are different because of the Mach Zehnder interferometer in Fig. \ref{schematic}, a comparison of the experimental results of a control group without turbulence and that of an experimental group with turbulence should be presented at first to point out the actual impact of extra perturbations. The experimental results of the relation between the SI and the overlapping degree of two Gaussian vortex beams are plotted in Fig. \ref{comparison}. Here, we selected a pair of coherent Gaussian vortex beams with opposite single charges and the width of each one at the receiving plane is $3.5\rm{mm}$. The displacement of BS3 is used as the abscissa and the position where two phase singularities coincide as the origin. Both the scintillation indexes at phase singularities and at the peaks of scintillation enhancement are given. The gray values of the pixels, which represent the mean intensities at phase singularities and at the peaks of scintillation enhancement, are presented correspondingly. By the way, the horizontal dashed lines denoting the unit gray value are plotted for convenient distinction. In order to guarantee the experimental accuracy, the realizations of each case were increased to $1000$.

It can be seen that there are significant differences in SI among all experimental groups. The large values of the scintillation indexes which are much larger than unity mainly caused by the inevitable noise of the CCD and the phase difference fluctuations of two optical vortices. The first factor would result in the fluctuations of the gray value, and that dominates the instability in the regions where the intensities should be zero. This is obviously reflected in the cases where the mean gray values are much smaller than unity in Fig. \ref{comparison}(a). Besides, in general, smaller mean gray value corresponds to larger value of SI. This phenomenon is more obvious in Fig. \ref{comparison}(a) rather than in Fig. \ref{comparison}(b). It indicates that the impact of extra perturbations gradually becomes larger in proportion to the impacts of other influencing factors with the increase of the strength of turbulence. The impact of the second factor varies over time, and we repeated data collection for each case as many times as possible to guarantee the impact was reduced to a minimum. Similarly, with the increase of the strength of turbulence, the impact becomes negligible compared to the impact of extra perturbations.

Comparing Fig. \ref{comparison}(a) with \ref{comparison}(b), for the groups whose gray values are much smaller than unity, the variations in the scintillation indexes at enhanced peak are wide. At the same time, the other groups whose gray values are around unity change slightly. Although the groups whose gray values are much smaller than unity are sensitive to the proportion increase of the impact of turbulence, the impact of the inevitable noise can not be ignored either. To verify the existence of scintillation enhancement, it should be demonstrated that this phenomenon is attributable to the extra perturbation. Therefore, only the experimental results whose gray values are larger than unity are suitable for further investigation. This conclusion can be supported by the comparison of the scintillation indexes at phase singularities as well. In Fig. \ref{comparison}(b), the larger the gray value is, the closer the SI at phase singularity is to the actual turbulence induced SI, namely around unity \cite{Aksenov:15}. Based on the above discussion, the groups whose displacements are around $0.6\rm{mm}$ or $-0.6\rm{mm}$ are the optimum choices. This state locates between the state where the independence of phase singularity begins to manifest and the state where the gray value at enhanced peak starts to drop to the value smaller than unity. Moreover, the state is in accordance with the state where scintillation enhancement appears \cite{wang2021phase}. In our experiment, we chose the state whose displacement is around $0.6\rm{mm}$ for further study.

\begin{figure}[ht!]
\centering\includegraphics[width=0.6\linewidth]{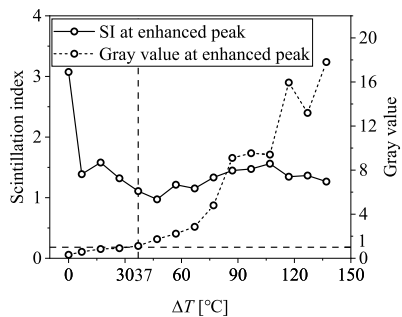}
\caption{Experimental results of the scintillation indexes and the gray values of the pixels at the peaks of scintillation enhancement versus $\Delta T$.}
\label{turbulencechange}
\end{figure}

After the combination of two Gaussian vortex beams is determined, the experimental results of the scintillation indexes and the gray values of the pixels at the peaks of scintillation enhancement versus $\Delta T$ are presented in Fig. \ref{turbulencechange}. The horizontal dashed line denotes unit gray value, and the vertical dashed line denotes $\Delta T=37^{\circ}\rm{C}$. The SI drops rapidly as the increase of $\Delta T$ when the mean gray value is smaller than unity (left side of the vertical dashed line). The noise of the CCD has great impact on the SI, and the results can not be applied to verification. On the right side of the vertical dashed line, the SI increases gradually until $\Delta T$ reaches $107^{\circ}\rm{C}$, and then begins to fall back. It is in agreement with the trend of that in Ref. \cite{wang2021phase}. But the variation range is narrower than the theoretical result. That is mainly caused by the different parameters of optical fields. In spite of the discrepancy, the gray value increasing along with the rise of $\Delta T$ supports the conclusion that the variation trend of SI is mainly caused by turbulence, and that further verify the existence of scintillation enhancement. 

\begin{figure}[ht!]
\centering\includegraphics[width=0.9\linewidth]{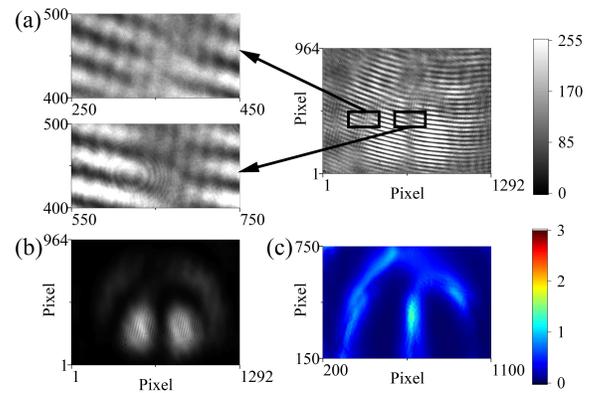}
\caption{Results of the experimental group of $\Delta T=107^{\circ}\rm{C}$ in Fig. \ref{turbulencechange}: (a) the interference pattern of the recombined beam and the reference beam, (b) the mean intensity distribution of the recombined beam and (c) the distribution of the SI.}
\label{fig5}
\end{figure}

So far, we have verified the existence of scintillation enhancement by experiment. The experiment makes it possible to figure out what exactly induces this phenomenon. We take the experimental group of $\Delta T=107^{\circ}\rm{C}$ in Fig. \ref{turbulencechange} for example. The interference pattern of the recombined beam and the reference beam, the mean intensity distribution of the recombined beam and the corresponding distribution of the scintillation index are presented in Fig. \ref{fig5}, respectively.

Two areas of the interference pattern (right side of fig. \ref{fig5}(a)), which are marked by black boxes, are magnified and shown on the left side of fig. \ref{fig5}(a). By means of the interference patterns in these areas, we only observe the presence of a screw dislocation on the left side. The screw dislocation on the right side integrates with an infinitely extended edge dislocation. The discrepancy leads to different distributions of SI in Fig. \ref{fig5}(c), namely the knot like structure on the left side and the peak in the middle. The former is induced by a screw dislocation, and the latter is caused by the combined effects of two types of phase discontinuities. 

\begin{figure}[ht!]
\centering\includegraphics[width=0.7\linewidth]{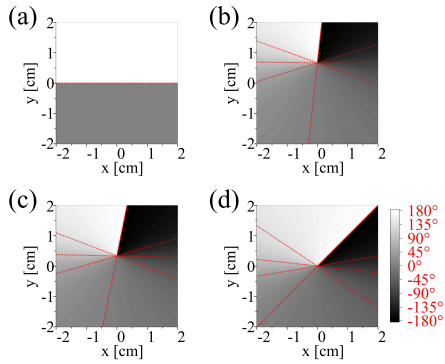}
\caption{Phase distributions of (a) only an infinitely extended edge dislocation existing and (b-d) both two types of phase dislocations existing, respectively.}
\label{fig6}
\end{figure}

However, the integration of two types of phase discontinuities does not mean coexistence. Here we present an example, namely an infinitely extended dislocation and a screw dislocation with different combinations, to illustrate the actual phase distribution. The phase maps of four combinations are shown in Fig. \ref{fig6}. The infinitely extended edge dislocation in each phase map is in line with $y = 0 \rm{cm}$. Then we add screw dislocations at $\left(0\rm{cm},0.2 \rm{cm}\right)$, $\left(0\rm{cm},0.1 \rm{cm}\right)$ and $\left(0\rm{cm},0 \rm{cm}\right)$ in Figs. \ref{fig6}(b-d), respectively. It is obvious that only an anisotropic dislocation \cite{FREUND1993247} exists in each figure after the screw dislocations are added. All the phase maps have contours of equal phase (red dashed lines) spaced by $45^{\circ}$. These contours indicate that the slope of the phase is steep near the x axis and gentle near the y axis always, although the locations of the phase singularities change with the variation of the locations of added screw dislocations. That anisotropy is the manifestation of the impact of infinitely extended edge dislocation on screw dislocation.

Return to Fig. \ref{fig5}, the singularity of the anisotropic dislocation overlaps with the enhanced peak, which further verifies that the scintillation enhancement is induced by the anisotropic dislocation rather than just a coincidence. Overall, we conjecture that the combined effects of two types of phase discontinuities, namely the effect of an anisotropic dislocation, induces scintillation enhancement. This phase structure results in great fluctuation of the intensity distribution of the optical field. To be more specific, the phase discontinuities lead to the intensity distribution shaped like a narrow valley in the middle of Fig. \ref{fig5}(b), and that results in the distribution of the SI shaped like a towering mountain. 

Based on the conjecture mentioned above, several experimental results are able to be explained. Firstly, the experimental results show that the scintillation enhancement occurs when two Gaussian vortex beams are partial overlapping, and the theoretical results in Ref. \cite{wang2021phase} predict this phenomenon fades away when they are complete overlapping. When two phase singularities are independent, but enter the influencing range of an infinitely extended edge dislocation, the impact of phase discontinuities shows the combined effect of screw dislocations and infinitely extended edge dislocations, namely the effects of two anisotropic dislocations. That makes the scintillation enhancement happen. As these two phase singularities get closer, the SI of the enhanced peak increases first and decreases latter. The annihilations of two anisotropic dislocations is the turning point of the variation of the SI. That means the scintillation enhancement in Ref. \cite{wang2021phase} is induced by two anisotropic dislocations. The different numbers of anisotropic dislocations may explain the discrepancy between the variations of the SI in Ref. \cite{wang2021phase} and in Fig. \ref{turbulencechange}. Secondly, how does phase discontinuities result in the trend of SI in Fig. \ref{turbulencechange} should be explained. The general expression of the SI of a beam under perturbations is described as follows \cite{2005Laser}
\begin{equation} \label{scintillation}
\sigma _I^2 = \frac{{\left\langle {{I^2}\left( {\boldsymbol{\rho },z} \right)} \right\rangle }}{{{{\bigl\langle {I\left( {\boldsymbol{\rho },z} \right)} \bigr\rangle }^2}}} - 1,
\end{equation}
where ${I\left( {\boldsymbol{\rho },z} \right)}$ and ${{I^2}\left( {\boldsymbol{\rho },z} \right)}$ denote the instantaneous intensity and the square of the instantaneous intensity at the plane $z$ with transverse coordinates $\boldsymbol{\rho} \equiv \left( {{\rho_x},{\rho_y}} \right)$. At the rising stage of SI, the high intensity area moves to the low intensity area occasionally, which is caused by extra perturbations. This random movement is known as beam wander \cite{2005Laser}. At this stage, the increase of the fluctuation of the intensity compared to the increase of the mean intensity is great enough to induce the rise of the SI. After these two processes reaching equilibrium, namely the increment of $\left\langle {{I^2}} \right\rangle$ equals to that of ${{\bigl\langle {I} \bigr\rangle }^2}$, the SI begins to drop gradually. 

In conclusion, we have verified the existence of scintillation enhancement by means of measuring the scintillation index of the recombined beam composed of two coherent Gaussian vortex beams with $\pm 1$ topological charges. By means of reference wave based interferometric method, we have verified that this phenomenon is induced by the combined effect of a screw dislocation and an infinitely extended edge dislocation, namely the impact of an anisotropic dislocation. Based on that, the theoretical results predicted in Ref. \cite{wang2021phase} have been explained. To the best of our knowledge, it is the first time to study the combined impact of two types of phase discontinuities on a laser beam under weak perturbations by experiment. The scintillation enhancement indicates that the anisotropic dislocation is more sensitive to perturbations than either type of phase discontinuity. That makes the anisotropic dislocation have potential for weak perturbation measurement. Our results may be useful in the researches of phase discontinuity and statistical optics.


\begin{backmatter}


\bmsection{Disclosures} The authors declare no conflicts of interest.

\bmsection{Data availability} Data underlying the results presented in this paper are not publicly available at this time but may be obtained from the authors upon reasonable request.

\end{backmatter}

\bibliography{sample}

\bibliographyfullrefs{sample}

\end{document}